\documentclass[11pt]{article}

\usepackage{amssymb,amsmath,textcomp}
\usepackage{graphicx,subfigure,cite}
\usepackage[linktocpage,unicode]{hyperref}

\def\esv{\aver{\overline{\Delta E}\sigma v}}
\newcommand{\aver}[1]{\left\langle \, #1 \, \right\rangle}

\begin{document}
\title{On the classical description of the recombination of dark matter particles with a Coulomb-like interaction}

\author{K.~M. Belotsky$^{1,2}$\thanks{k-belotsky@yandex.ru}, E.~A.~Esipova$^{1}$\thanks{esipovaea@gmail.com}, A.~A. Kirillov$^{1,3}$\thanks{kirillov-aa@yandex.ru}\\
$^1$ National Research Nuclear University MEPhI\\ (Moscow Engineering Physics Institute), Moscow, Russia\\
$^2$ Centre for Cosmoparticle Physics ``Cosmion'', Moscow, Russia\\
$^3$ Yaroslavl State P.~G.~Demidov University, Yaroslavl, Russia}

\date{}

\maketitle

\begin{abstract}
Cold dark matter (DM) scenario may be cured of several problems by involving self-interaction of dark matter.
Viability of the models of long-range interacting DM crucially depends on the effectiveness of recombination of the DM particles, making thereby their interaction short-range. Usually in numeric calculations, recombination is described by cross section obtained on a feasible quantum level. However in a wide range of parameter values, a classical treatment, where the particles are bound due to dipole radiation,
is applicable. The cross sections, obtained in both approaches, are very different and lead to diverse consequences. Classical cross section has a steeper dependence on relative velocity, what leads to the fact that, after decoupling of DM particles from thermal background of ``dark photons'' (carriers of DM long-range interaction), recombination process does not ``freeze out'', diminishing gradually density of unbound DM particles.
Our simplified estimates show, that at the taken parameter values (the mass of DM particle is $100$ GeV, interaction constant is $100^{-1}$, and quite natural assumptions on initial conditions, from which the result is very weakly dependent) the difference in residual density reaches about $5$ orders of magnitude on pre-galactic stage. This estimate takes into account thermal effects induced by dipole radiation and recombination, which
resulted in the increase of both temperature and density of DM particles by a half order of magnitude.
\end{abstract}

  
The models of self-interacting dark matter (DM) have aroused a lot of interest in the last time \cite{2010JCAP...05..021K, 2011MPLA...26.2823K, 2012arXiv1212.6087B, 2014JCAP...07..039P, 2014PhRvD..90e5030W, 2014JCAP...12..033V, 2014arXiv1403.1212B, 2015PhRvD..91h3532P, 2015arXiv150500109P, 2015PhRvD..91b3512F, 2015PhRvD..91d3519K}. DM with long-range interaction (referring hereafter as to $y$-interaction) seems to be able to escape several problems of ordinary cold dark matter (CDM) scenario, such as an overproduction of subhalos and cuspy density profile in them \cite{2013PhRvD..87j3515C, 2014PhRvD..90d3524B, 2015PhRvD..91h3532P, 2015PhRvD..91b3512F}. At the same time, an ellipticity of big halos is not spoiled at some model parameters \cite{2015PhRvD..91b3512F}.  
An enhancement of annihilation signal in the Galaxy (so called Sommerfeld-Gamov-Sakharov enhancement \cite{1928ZPhy...51..204G, 1931AnP...403..257S, 1948ZETF...18..631}), considered for the first time (to our knowledge) in \cite{2000GrCo....6..140}, is one more possible bonus of the models of question.
Analysis of recent observations of forming galactic cluster Abell 3827 also favours self-interacting DM \cite{2015MNRAS.449.3393M}.
Origin of supermassive black holes can be connected with an existence of DM component with strong self-interaction \cite{arxiv.1501.00017}. Generally, models with dissipative form of DM as sub-component find more applications \cite{arxiv.1312.1336v2, Randall}.

Essential feature of cosmological evolution of $y$-interacting DM is a formation of atomic-like bound states by DM particles with opposite $y$-charges. If oppositely $y$-charged particles are particle ($a$) and anti-particle ($b=\bar a$), then they annihilate, what may drastically affect their residual density \cite{2005GrCo...11...27B, 2014JCAP...12..033V, 2013PhRvD..87j3515C}. If the bound particles are different species ($a$ and $b\ne \bar{a}$) so bound state is stable, then depending on relative amount of bound and unbound particles, as it is obtained by the period of large scale structure formation, DM dynamics is very different and whether or not the model gets constraint from observations \cite{2013PhRvD..87j3515C, 2014PhRvD..90d3524B, 2014JCAP...07..039P}. Description of recombination process plays a clue role here. 
Usually a quantum approach is used for it. However a classical approach, which 
was used, in particular, for magnetic monopoles and heavy neutrinos \cite{1978PhLB...79..239Z, 2005GrCo...11...27B}, seems to be valid in a broad interval of parameter values. It leads to a result very different from that obtained on quantum level, which in the commonly accepted form does not come to the classical limit.

Classical recombination cross section is obtained from condition that the scattered particles lose, due to dipole radiation, sufficient energy to get bound \cite{LandauBook-2, 1978TMP....34..112E}, and is given by
\begin{equation}
	\sigma_{\rm rec}=\pi \rho_{\max}^2(v)=(4\pi)^{2/5}\pi\,\frac{\alpha_y^2}{\mu^2}\frac{1}{v^{14/5}},
	\label{scl}
\end{equation}
where $\rho_{\max}$ is the maximal impact parameter at which a pair is bound, $\alpha_y$ is the constant of $y$-interaction, $\mu=\frac{m_am_b}{m_a+m_b}$ is the reduced mass of the pair of the scattered particles with $m_a$ and $m_b$ being their masses ($m_a\le m_b$), $v$ is their initial relative velocity. This cross section has a steeper dependence on velocity with respect to that of usually accepted quantum 
recombination cross section. 
The latter in form of Kramers' formula \cite{Kramers} summed over all quantum levels is (valid for $v\ll \alpha_y$):
\begin{equation}
\sigma_{\rm rec}=\frac{32\pi}{3\sqrt{3}}\frac{\alpha_y^3}{\mu^2}\frac{\ln (v^{-1})}{v^2}.
\label{sq}
\end{equation}

Steeper behaviour of classical cross section leads to the fact that recombination process does not freeze out on both radiation dominated (RD) and (even faster expanding) matter dominated (MD) stages, and relative number of unbound $y$-charged particles falls down gradually with time.

Classical formula \eqref{scl} is assumed to be valid when \cite{1978TMP....34..112E}
\begin{equation}
v\ll \alpha_y^{5/2}.
\label{cl}
\end{equation}
Under this condition, binding is found to occur predominantly due to multiple soft photon emission, what allows classical treatment. However considering on quantum level, only one-, two-photon final states are usually taken into account. Eq.\eqref{cl} can be formally deduced from condition that binding of two particles (i.e.\ when initial kinetic energy of relative motion, $E_{\rm rel}=\mu v^2/2$, is lost) happens at distance ($R_{\rm b}$) much greater than the radius of the respective ground bound state ($a_{\rm B}$). In this case, action of the system, as will be shown, becomes much greater 1 (in units $\hbar=1$) and thus the classical approach is reasonable.

Below we shortly discuss classical approach implications in recombining DM cosmological evolution. It includes estimations, some of which may seem to be simple, to trace explicit dependence of result on the parameters.

\bigskip

Our results will basically relate to parameter region lying around fiducial values $m_a\sim \mu=100$ GeV and $\alpha_y=1/100$ used for numerical estimations.
Also for definiteness we assume (as in case of heavy neutrino model \cite{2005GrCo...11...27B}) that before a direct annihilation of $a$ and $b$ (where either $b=\bar{a}$ or $b\ne \bar{a}$), happening when the temperature becomes below their mass (of the lightest), $T=T_*\sim m_a/10$, $y$-plasma (consisting of $a$, $b$ with their antiparticles and $y$-photons - massless $y$-force carriers) has the same temperature as ordinary matter. Right after annihilation, $y$-photon background decouples from $a$ and $b$ as well as from ordinary matter ($O$), while the opposite is not true. That is the $y$-background is no longer influenced by $a$ and $b$, and by $O$ through $O$-$a,b$ possible coupling, but $a$ and $b$ are influenced by $y$. Starting from this moment, the temperature of $y$-background ($T_y$), as of a closed system, changes as inverse scale factor, whereas that of ordinary photons feels also entropy re-distribution between ordinary matter components. So for $y$- and $O$- photons' temperature relation one has
\begin{equation}
T_y=\kappa^{1/3}T,\qquad \kappa(T)=\frac{g_{s,o}(T)}{g_{s,o}(T_*)},
\end{equation}
where $g_{s,o}(T)$ is the effective number of ordinary matter species (excluding $y$) contributing into entropy density ($s$). For the chosen numeric values, contribution of $y$ in density at nucleosynthesis (BBN) makes up $\kappa^{4/3}(T\sim 1\, {\rm MeV})\approx 0.06$ from that of $O$-photons, what has no effect on BBN data. The temperature of $a$ and $b$ is equal to $T_y$ until they decouple from $y$, influence of $O$-matter (being determined by some weak scale interaction) is negligible (see Appendix A). After decoupling it evolves as (see Appendix A)
\begin{eqnarray}
T_a\approx T^2/\bar{T}_{ay},\qquad 
\bar{T}_{ay}=\frac{\pi^{3/4}g_{\epsilon}^{1/4}m_a^{3/2}}{2^{5/2}5^{1/4}\zeta(3)^{1/2}m^{1/2}_{\rm Pl}\kappa\alpha_y}\approx 0.2\;{\rm MeV} \left(\frac{m_a}{100\;{\rm GeV}}\right)^{3/2}\frac{1/100}{\alpha_y}.
\label{Ta}
\end{eqnarray}
Here $g_{\epsilon}$, being effective number of matter species (including $y$) contributing into energy density, as well as $\kappa$ are taken at $T=0.2$ MeV.

Evolution of number density of unbound $a$- and $b$- species can be approximately described by equation system (see Appendix B)
\begin{equation}
	\begin{cases}
		\cfrac{dr}{dT}=\aver{\sigma_{\rm rec} v}\cfrac{r^2s}{HT}\\
		\cfrac{d\theta}{dT}=-\cfrac{\bar{T}_{ay}}{T^2}\aver{\left(T_a-\cfrac{1}{3}\; E_{\rm pair}-\cfrac{2}{9}\; E_{\rm rel}\right)\sigma_{\rm rec}v} \cfrac{rs}{HT}.
	\end{cases}
\label{system}
\end{equation}
Here $r$ is the number density conventionally expressed in units of $s$, the brackets ``$\aver{}$'' mean averaging over velocity distribution of $a$ and $b$, $\theta$ shows deviation of $T_a$ from Eq.(\ref{Ta}), $E_{\rm pair}=E_a+E_b$ is the energy lost by $a$-,$b$-gas (thermal bath) in the result of the pair binding (that is the case of annihilation, i.e. when $b=\bar{a}$, or when the bound systems are thermally decoupled from $a$-, $b$-gas). The second equation takes into account thermal effects, caused by scattering of particles: 
(*) presumably slower particles to be bound ``go out'' (annihilate or decouple) of $a$-,$b$-plasma, effectively heating it;
(**) scattered but unbound pairs experience dipole energy losses cooling plasma. 
Evolution is considered in terms of $O$-photon temperature ($T$).

Here we do not take into account some recombination process details which are more appropriate for quantum case (such as recombination into different level bound states, 
red-shifting of recombination photons), as well as inverse processes (which are not important for the big parameter space of question), and also in quantum case the second equation of \eqref{system} is omitted ($\theta\equiv 1$).

The bound states start to form when $T_a$ becomes much lower than ionization potential $I=\mu \alpha_y^2/2$, $T_a<T_{a\,{\rm rec}}\sim I/10$. For the chosen values, it is $T_{a\,{\rm rec}}=0.5$ MeV, at which $T_a=T_y$, what corresponds to $O$-matter temperature $T=T_{\rm rec}=\kappa^{-1/3} T_{a\,{\rm rec}}\approx 1$ MeV. 
Depending on behaviour of $\aver{\sigma_{\rm rec} v}$ with $T$ and epoch, recombination process should flow in different regimes. Basically, it either damps (freezes out) and it is initial moment (when it starts) what predetermines the residual density of free $a$ and $b$, or it ``burns'' continually with a self-adjusted rate and final moment, at which we need to know the density, defines its value.
For classical cross section at $T\sim T_{\rm rec}$ recombination process gives effect and temporarily freezes out, but after $a$-$y$ decoupling it is restored and goes with a steady rate until the galactic stage\footnote{The moment $T_a\sim I/10$ can come after $a$-$y$ decoupling at some parameter values and the first freezing stage is absent.}. Herewith, recombination rate $\Gamma_{\rm rec}=n\aver{\sigma_{\rm rec} v}$ 
turns out to be of the same order as Hubble rate ($H$), if thermal effects (second equation of system \eqref{system}) are ignored ($\theta \equiv 1$), otherwise ($\theta \ne 1$) the ratio $\Gamma_{\rm rec}/H$ slowly goes down with time. So the value of residual density is not sensitive to the initial moment and all early history, but fully defined by the final one, and is just weakly dependent on initial abundance for $\theta\ne 1$ (see below).

Classical approach is assumed to be valid when $R_b$ as well as energy loss length-scale\footnote{It is defined as the length on which 90\% of initial energy is lost due to dipole radiation before the particles are bound.}, $L_{\rm loss}$, are much greater than dark atomic size $a_B$. From other side, they must be much less than spacing between $y$-charged particles, $L_{\rm sp}$ (\textit{a fortiori} $a$-$a(b),y$ interaction lengths). $R_b$ and $L_{\rm loss}$ can be found by taking mechanical energy (kinetic plus potential in cms) loss rate equal to dipole radiation intensity
$$\frac{dE}{dt}=-\frac{1}{6\pi}\ddot{\mathbf{d}}^2.$$
Dipole moment can be expressed with the help of Newton's law, $\mathbf{d}=e_y\ddot{\mathbf{r}}$, $|\ddot{\mathbf{r}}|=\frac{\alpha_y}{r^2} \frac{1}{\mu}$ (braking due to radiation is negligible here), and then through relation $d t=d r/\left[\frac{2}{\mu} \left( E+ \frac{\alpha_y}{r} -\frac{M^2}{2\mu r^2} \right)\right]^{1/2}$ one comes to equation for $E$ from distance between the particles $r$
\begin{equation}
	\frac{d E}{dr} = -\frac{2}{3} \frac{\alpha_y^3}{\mu^2} \frac{1}{r^4} \frac{1}{\sqrt{\cfrac{2}{\mu} \left( E+ \cfrac{\alpha_y}{r} -\cfrac{M^2}{2\mu r^2} \right)}}.
	\label {eq2}
\end{equation}
Here $M$ is the angular momentum, which can be assumed to be conserved. In the region of interest, solution of Eq.\eqref{eq2} can be simplified by neglecting ``$E$'' in the square root.

\begin{figure}[t]
 \centering
 \subfigure[\label{Rbrho}]{\includegraphics[width=.45\textwidth]{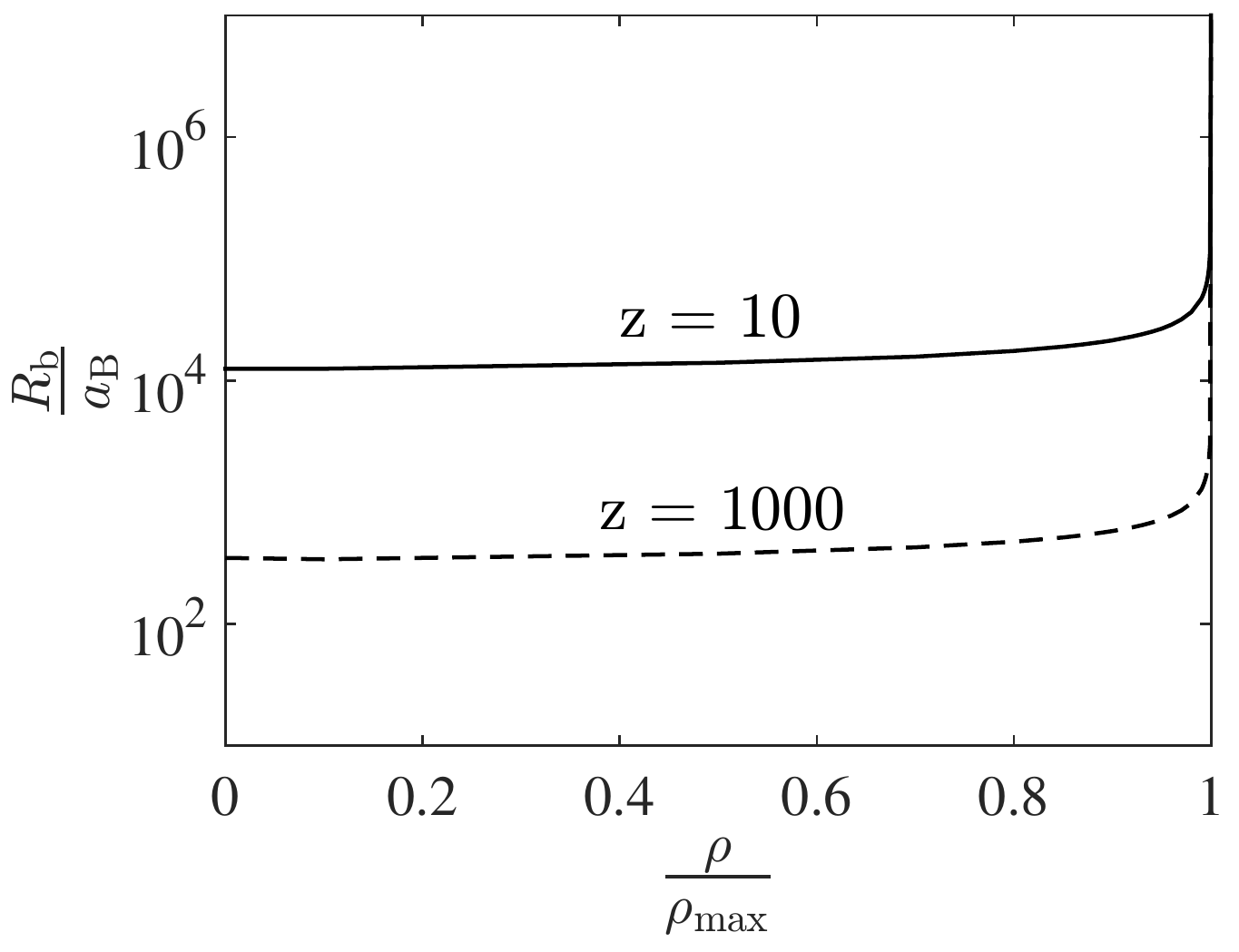}}
 \qquad
 \subfigure[\label{RbT}]{\includegraphics[width=.45\textwidth]{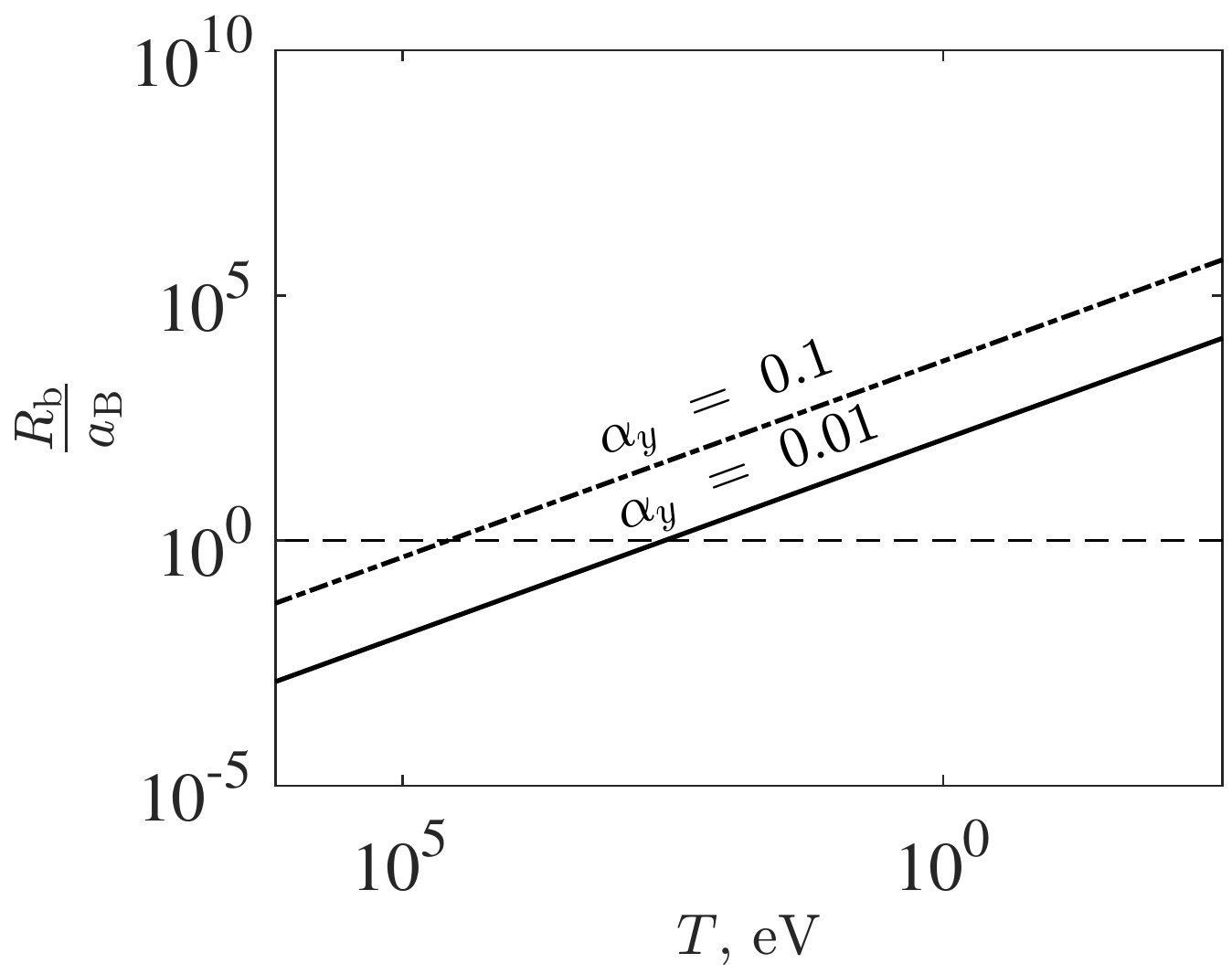}}
 \caption{The ratio $R_{b}/a_{B}$ as function of impact parameter for two redshift values (left) and of temperature for two values of $y$-interaction constant (right).}
\end{figure}

Condition $R_b,L_{\rm loss}\ll L_{\rm sp}$ is found to be true by a wide margin for the most of parameter values at any epoch of question \cite{2015mephi.conf.}. 
The values $R_b$ and $L_{\rm loss}$ are of the same order of magnitude and both depend on impact parameter $\rho$ and on energy, which falls down with the Universe expansion. Dependences of $R_b/a_B$ from $\rho/\rho_{\rm max}$ and from the temperature of $O$-photon are shown in Figs~(\ref{Rbrho}, \ref{RbT}). One can see that $R_b/a_B$ at the most of $\rho$ values is close to that at $\rho=0$, therefore $R_b/a_B(\rho=0)$ is used in all further calculations. In this case $R_b\approx \frac{2}{v^{4/5}}\frac{\alpha_y}{\mu}$ and $E\approx \frac{\mu v^2}{2}\left[1 - \left( \frac{R_b}{r} \right)^{5/2} \right]$.

Since the late period is determinative for residual density value in case of classical cross section, we single out late expansion moment, when the galaxies start to form, $z\sim 10$ ($T\sim 30$ K). As one can see from Fig~(\ref{RbT}), $R_b/a_B\gg 1$ there. This ratio depends on parameters as $R_b/a_B\propto \alpha_y^2 \mu^{2/5}/T_a^{2/5}\sim$(after $a$-$y$ decoupling)$\sim\alpha_y^{8/5}\mu/T^{4/5}$ (for $\theta=1$). At the chosen parameter values, classical approximation does not work for $T\gtrsim T_\text{q-c}\approx 200$ eV, but, importantly, it does in the late period.

Solution of the system (\ref{system}) is given by Eqs.\eqref{rtheta}--\eqref{rapprox} of Appendix B. In the Fig.(\ref{r}) the density of unbound $y$-charged particles, obtained in different approximations, is shown as function of $T$. In classical case, the mentioned above features are seen: sharp recombination effect freezes out around $T\sim 1$ MeV, continual recombination takes place from $T\sim 0.1$ MeV until the late epoch, changing slope at RD$\rightarrow$MD transition. (All sharp breaks in the curve would be smoothed if estimated more accurately.) Quantum formula gives very small effect at the chosen parameters: there are unessential effect at $T\sim 1$ MeV and a slow logarithmic decline of $r(T)$ from $y$-$a$ decoupling to RD$\rightarrow$MD transition. The curve ``Q-C'' in the figure is obtained with application of Eq.\eqref{sq} in interval $T>T_\text{q-c}$ and Eq.\eqref{scl} at $T<T_\text{q-c}$. As one can see, in the late period this solution comes to that obtained in pure classical approximation. A small deviation is caused by thermal effects (described by second equation of the system \eqref{system}) which make the result being a little sensitive to the initial density (see Eq.\eqref{rapprox} of Appendix~B). In calculations the following values had been used: $\alpha_y=1/100$, $m_a=100$ GeV, $m_b=1$ TeV, $r(T_{\rm rec})=\rho_{\rm CDM}/m_b/s(T=2.7\,{\rm K})\approx 4.6\times 10^{-13}$ with $\rho_{\rm CDM}\approx 1.4$ keV$/$cm$^3$ being the modern CDM density. Also, quantum recombination rate was taken from \cite{2010JCAP...05..021K}, where $\sigma_{\rm rec}$  is a little corrected as compared to Eq.\eqref{sq}.


\begin{figure}[t]
	\centering
	\subfigure[]{\includegraphics[width=.50\textwidth]{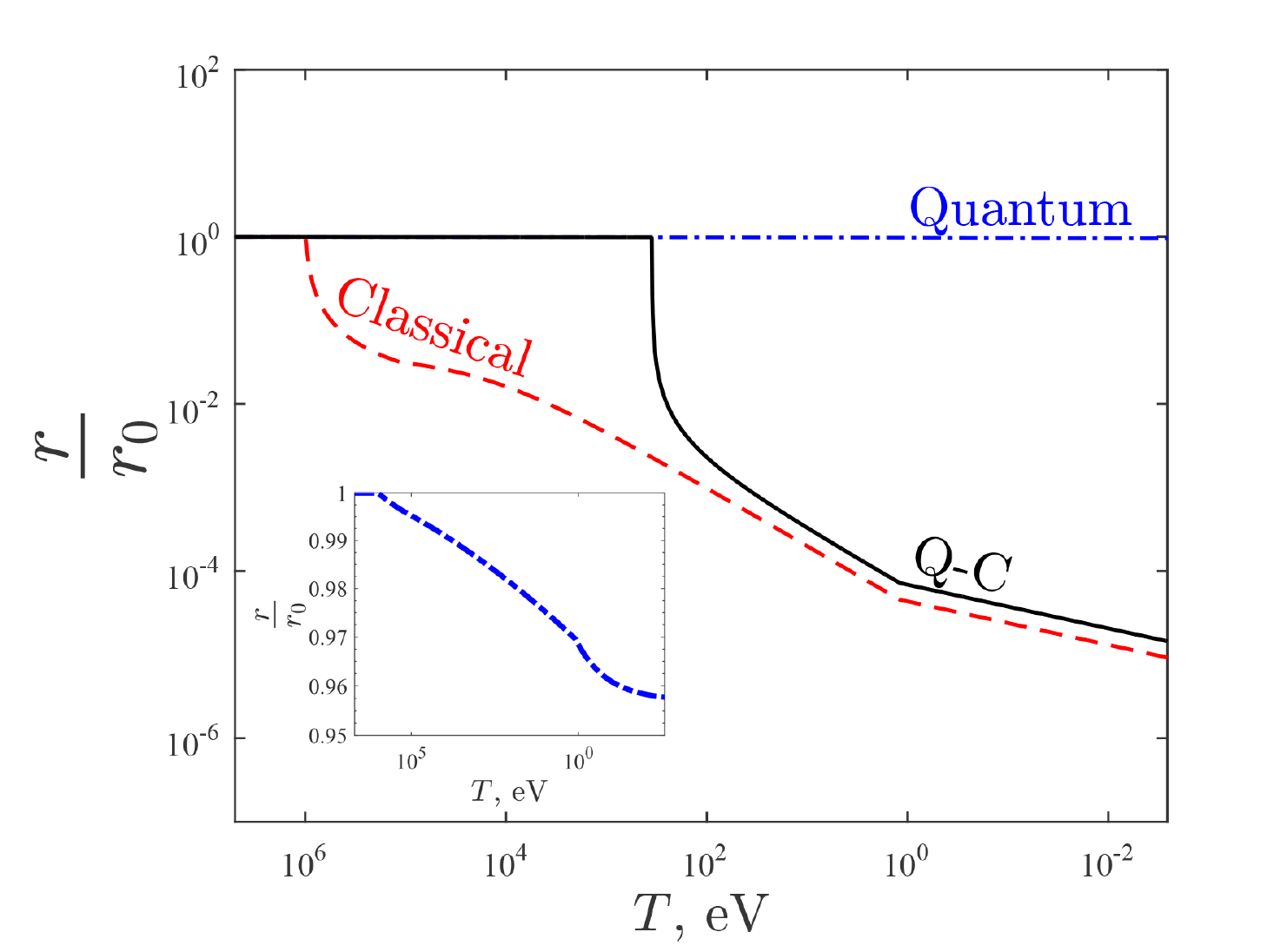}\label{r}}
	\qquad
	\subfigure[]{\includegraphics[width=.43\textwidth]{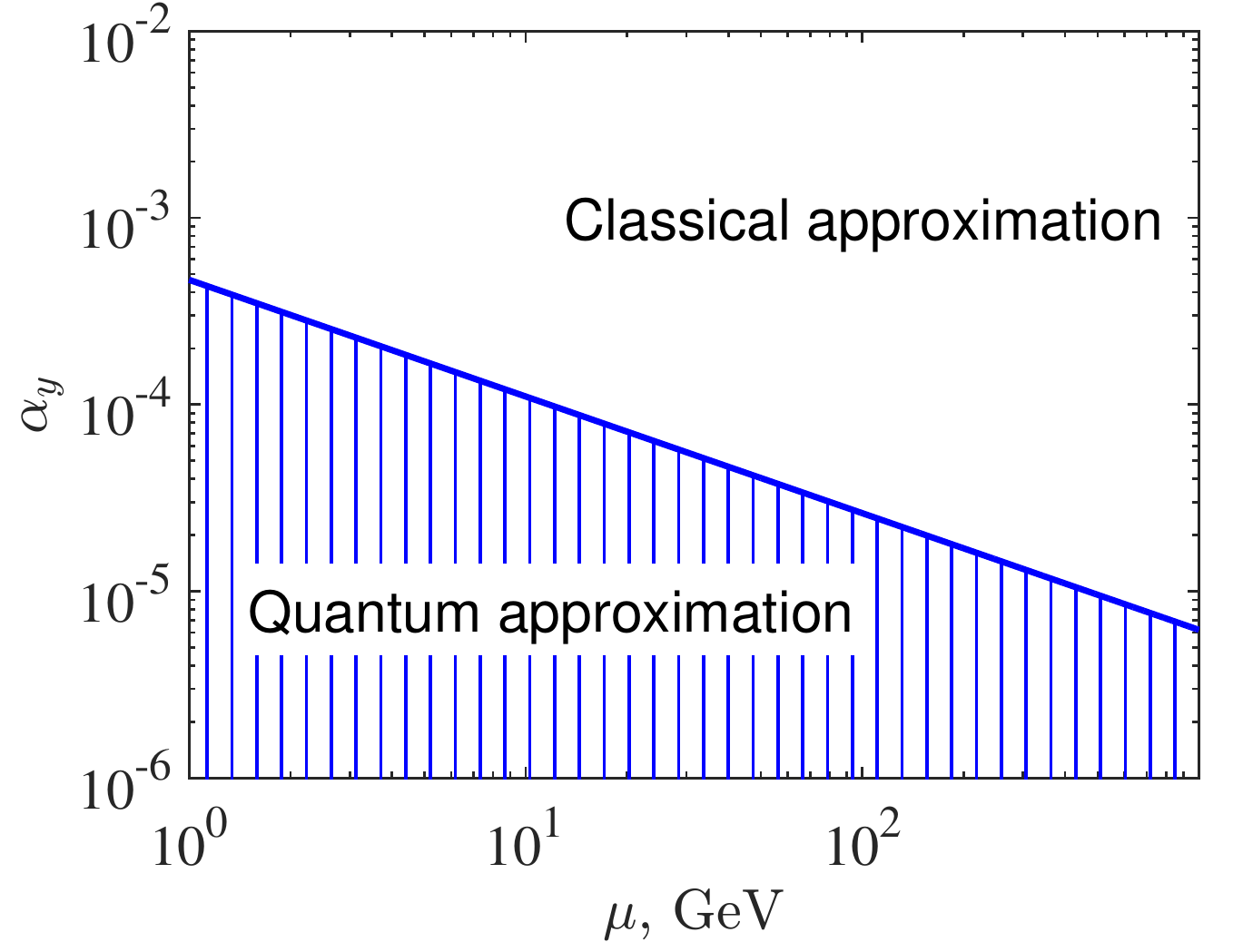}\label{reg}}
	\caption{\textit{Left}: relative density of unbound $y$-charged DM particles as a function of $T$, obtained with either classical (dashed curve) or quantum (dashed-dot curve) recombination cross sections ($\sigma_{\rm rec}$). The curve ``Q-C'' is obtained applying before and after the moment $T\approx 200$ eV quantum and classical $\sigma_{\rm rec}$ respectively. On the small plot inside, a quantum case is shown with the scaled vertical axes. \textit{Right}: the region of values $\mu$ and $\alpha_y$, where classical cross section should be used for estimation of residual density of (un)bound DM particles.}
\end{figure}

Note that the mentioned above thermal effects in classical case 
are found to be weak ($\theta(T)\sim T^{-1/23}$), but nonetheless noticeable. At $T\sim 30$ K for $\mu=100$ GeV and $\alpha_y=1/100$,  $\theta\approx 4$. The ratio of $r(T\sim 30)$, obtained with and without account for thermal effects (using only classical cross section), is $\sim 3$.

Condition \eqref{cl} for period $z\sim 10$ allows to outline the region of parameters values ($\mu$ and $\alpha_y$) when the classical approximation for $\sigma_{\rm rec}$ is applicable. It is shown in the Fig.(\ref{reg}). There can exist a region of parameter space where calculations should be re-considered, since interaction radius $\sim \alpha_y/T_a$ exceeds spacing between DM particles \cite{2015mephi.conf.}.

Finally, we show arguments in favour of condition Eq.\eqref{cl} on the base of action in the Appendix~C.

\vspace{0.5cm}

To conclude, we have shown that classical recombination cross section Eq.(\ref{scl}) can be applicable in a broad parameter region of self-interacting DM models, and it leads to strongly different result comparing to obtained with usually accepted quantum cross section. It may change (extend) the parameter region of the corresponding models' viability.

\subsection*{Acknowledgements}

We would like to thank M.~Yu.~Khlopov, cooperative work with whom initiated this paper, S.~G.~Rubin and A.~E.~Dmitriev for interest and useful discussions, M.~N.~Laletin for his help in writing the text. 

This work was performed within the framework of Fundamental Interactions and Particle Physics Research Center supported by MEPhI Academic Excellence Project (contract \textnumero~02.a03.21.0005, 27.08.2013), grant of RFBR (\textnumero 14-22-03048) and in part by grants of Ministry of Education and Science of the Russian Federation (\textnumero~3.472.2014/K) and the Russian Science Foundation (\textnumero~15-12-10039).

\subsection*{Appendix A}

Here we trace shortly the temperature evolution of $y$-charged particles, $a$ ($b$), before recombination starts. They are assumed to experience energy exchange with $y$-background and, possibly, ordinary matter. In calculations we suppose that $a$ interacts with $O$ as Dirac heavy neutrino.

To find $T_a$ one can formally use the first law of thermodynamics, which for $a$-particles can be reduced to
\begin{equation}
\frac{3}{2}\frac{dT_a}{dt}=\esv_{ay}n_y+\esv_{ao}n_o-3T_aH,
\label{dT}
\end{equation}
where $o=e,\nu,p,n,...$ are available species of $O$-matter, $\esv_{ai}$ is the kinematically averaged energy transfer in $ai$-interaction multiplied by respective cross section and relative velocity, averaged in thermal distribution, $n_i$ is the respective number density ($i=y$ or $o$). 

In all scattering processes of interest $a$ is non-relativistic, $i$ is ultra-relativistic ($p, n$ turn out to be too suppressed in density, so play no role in heat transfer with $a$). We adopt Boltzmann approximation for all species distributions. The $ay$-scattering is well described by Thomson cross section over the great part of parameter space. Then calculation of $\esv_{ay}$ with accuracy $\sim T/m_a$ gives
\begin{equation}
\esv_{ay}\approx \sigma_T \left( \frac{\aver{\omega^2}}{m_a}-\frac{8\aver{\omega}\aver{E}}{3m_a} \right)=\frac{32\pi \alpha_y^2}{m_a^3}T_y(T_y-T_a),
\label{esv2}
\end{equation}
where $\omega$ and $E$ are the energies of $y$- and $a$-particles respectively.

To estimate $\esv_{ao}$, assuming that $a$ is heavy neutrino, one needs to take the cross sections of relevant processes ($a\nu$, $a\bar{\nu}$, $ae^-$, $ae^+$), which in our limit are undistinguishable for particle and antiparticle  
\begin{equation}
\sigma_{a\nu}=\sigma_{a\bar{\nu}}=\frac{G_F^2\omega_{\rm lab}^2}{2\pi},\quad
\sigma_{ae^-}=\sigma_{ae^+}=\frac{G_F^2\xi_e\omega_{\rm lab}^2}{2\pi}.
\end{equation}
Here $\omega_{\rm lab}$ is the energy of incident $i$-particle in the reference frame where $a$ is in the rest, $G_F$ is the Fermi constant, $\xi_e=1-4\xi+8\xi^2\approx 0.50$ with $\xi=\sin^2\theta_W$ being the weak mixing parameter.
Unlike $ay$-scattering, $ae$-, $a\nu$- cross sections depend on energy, what accounts for higher power of the temperature in the final expression
\begin{eqnarray}
\esv_{a\nu(\bar\nu)}=\frac{180G_F^2}{\pi m_a}T_{\nu}^3(T_{\nu}-T_a),\quad
\esv_{ae^{\pm}}=\frac{180G_F^2\xi_e}{\pi m_a}T_e^3(T_e-T_a).
\label{esvNnue}
\end{eqnarray}

Eq.\eqref{dT} can be then re-written as
\begin{equation}
\frac{1}{2}\frac{dT_a}{dT}=-\frac{T(\kappa^{1/3}T-T_a)}{T_{ay}^2}-\frac{T^3(T-T_a)}{T_{ao}^4}+\frac{T_a}{T}.
\label{dT2}
\end{equation}
$T_{\nu}=T_e=T$ was put. Coefficients $T_{ay}\sim 0.1$ MeV, $T_{ao}\sim 10$ MeV. Second term in the r.h. of Eq.\eqref{dT2} has no effect on solution.
Excluding it, solution can be expressed in the form
\begin{equation}
T_a(T)=\frac{\sqrt{\pi}\kappa^{1/3}T^2}{T_{ay}}\exp{(T^2/T_{ay}^2)}(1-{\rm erf}(T/T_{Ny})).
\label{Ta1}
\end{equation}
At $T\ll T_{ay}$ (i.e. after $a$-$y$ decoupling)
\begin{equation}
T_a\approx \frac{T^2}{\bar{T}_{ay}},\quad \bar{T}_{ay}=\frac{T_{ay}}{\sqrt{\pi}\kappa^{1/3}}=\frac{\pi^{3/4}g_{\epsilon}^{1/4}m_a^{3/2}}{2^{5/2}5^{1/4}\zeta(3)^{1/2}m^{1/2}_{\rm Pl}\kappa\alpha_y}.
\label{Ta2}
\end{equation}

\subsection*{Appendix B}

Evolution of abundance of free $a$-, $b$- particles can be described by Boltzmann equation
\begin{equation}
\frac{dn}{dt}=-n^2\aver{\sigma_{\rm rec} v}-3Hn,
\label{dndt}
\end{equation}
which is easily reduced to the first equation of the system Eq.~(\ref{system}) with the help of replacements: $n=rs=r\frac{2\pi^2g_{s}}{45}T^3$, $-dt=\frac{1}{H}\frac{dT}{T}$ (for $g_{\epsilon,s}\approx \text{const}$). For calculation one parametrizes $\sigma_{\rm rec}=\sigma_0/v^{\beta}$. For $\aver{\sigma_{\rm rec} v}$ then one can get 
\begin{equation}
	\aver{\sigma_{\rm rec} v}=\frac{2\, \Gamma\left(2-\frac{\beta}{2}\right)}{2^{\tfrac{\beta-1}{2}}\sqrt{\pi}}\sigma_0
	\left(\frac{\mu}{T_a}\right)^{\tfrac{\beta-1}{2}}.
	\label{Grec}
\end{equation}

Since recombination rate is strongly temperature dependent (especially in classical case), thermal effects of recombination process itself can be important, correcting temperature evolution. 
These effects are relevant, obviously, after $a$-$y$ decoupling.

One can take again the first law of thermodynamics, $dQ=\delta A+dU$. One has the total particle number in some volume $N_{ab}=n_{ab}V$ with $n_{ab}=n_a+n_{b}=2n$ being their number density, the pressure $p=n_{ab}T_a$. 
Expansion of the Universe is treated as a work of gas: 
$\delta A=pdV=n_{ab}T_a\, 3HVdt$. Inner energy gain is $dU=\frac{3}{2}d(N_{ab}T_a)=\frac{3}{2}N_{ab}dT_a+\frac{3}{2}T_adN_{ab}$. Here we assume that $dN_{ab}=-2\aver{\sigma_{\rm rec} v}n^2Vdt$ as if the recombined pairs disappear from $a$-, $b$- gas. It would be true when $b=\bar a$, and also when $b\ne \bar a$ if the bound systems are out of thermal equilibrium with free $a$ and $b$, however it is not always the case \cite{2013PhRvD..87j3515C}. We do not see the opposite since it is not of principle here.

One can define $dQ$ through the energy lost during $ab$-scattering. If the pair is combined (impact parameter $\rho<\rho_{\rm max}$), then their energy $E_{\rm pair}=E_a+E_b=\frac{m_av_a^2}{2}+\frac{m_bv_b^2}{2}$ is lost completely. Otherwise ($\rho>\rho_{\rm max}$), their energy is lost partially due to dipole radiation, what is for given $\rho$ and
$v=|\vec{v}_a-\vec{v}_b|$ (in large scattering angle limit, which is realized when $\rho\ll \alpha_y/(\mu v^2)$) \cite{LandauBook-2}
\footnote{$aa$-, $bb$-scatterings do not lead to dipole radiation}
\begin{equation}
\Delta E(v,\rho)=\frac{2\pi\alpha_y^5}{\mu^4v^5\rho^5}=E_{\rm rel}
\left(\frac{\rho_{\max}}{\rho}\right)^5,
\label{DE}
\end{equation}
where $E_{\rm rel}=\mu v^2/2$ is the energy of relative motion. So, energy losses rate by $ab$-gas per unit volume can be given by
\begin{multline}
	\label{epslost}
	\dot{\varepsilon}=n_an_b\int \left\{ \int_0^{\rho_{\rm max}}(E_a+E_b)v2\pi\rho \,d\rho+
	\int_{\rho_{\rm max}}^D\Delta Ev2\pi\rho \,d\rho \right\}
	f_a(\vec{v}_a)d^3v_a f_b(\vec{v}_b)\,d^3v_b=\\ 
	=n^2\aver{E_{\rm pair}\sigma_{\rm rec}v}+\frac{2}{3}n^2\aver{E_{\rm rel}\sigma_{\rm rec} v},
\end{multline}
where $f_{a,b}$ is the distribution in velocity (Maxwell).
Upper limit $D$ should be given by Debye length of $ab$-plasma, but thanks to fast convergence of $\Delta E(v,\rho)$ with $\rho\rightarrow \infty$, we put 
$D\rightarrow \infty$ in Eq.(\ref{epslost}). 
Averaging over velocity distributions gives 
\begin{equation}
	\aver{E_{\rm pair}\sigma_\text{rec}v}=\frac{(7-\beta)\Gamma(2-\frac{\beta}{2})}{2^{\frac{\beta-1}{2}}\sqrt{\pi}}\sigma_0
	\left(\frac{\mu}{T_a}\right)^{\frac{\beta-1}{2}}T_a,
	\label{esbv}
\end{equation}
\begin{equation}
\aver{E_{\rm rel}\sigma_\text{rec} v}=\frac{(4-\beta)\Gamma(2-\frac{\beta}{2})}{2^{\frac{\beta-1}{2}}\sqrt{\pi}}\sigma_0
\left(\frac{\mu}{T_a}\right)^{\frac{\beta-1}{2}}T_a.
\label{epsdip}
\end{equation}

Combining $dQ=-\dot \varepsilon Vdt$ with other terms of the 1st law of thermodynamics gives
\begin{equation}
\frac{3}{2}\frac{dT_a}{dt}=\aver{\left(\frac{3}{2}T_a-\frac{1}{2}E_{\rm pair}-\frac{1}{3}E_{\rm rel}\right)\sigma_{\rm rec}v}n-3T_aH.
\label{dT3}
\end{equation}
Note, that the first two terms in the right side of Eq.(\ref{dT3}), $\propto \frac{3}{2}T_N-\frac{1}{2}E_{\rm pair}\propto (\beta-1)$, originate from the fact of disappearance (recombination) of the pair and lead to a heating of $ab$-gas (at $\beta>1$). Term $\propto \frac{1}{3}E_{\rm rel}$ does not dominate 
for $\sigma_{\rm rec}$ given by Eq.(\ref{scl}) and diminishes this effect. It is clear after accounting for Eqs.\eqref{Grec},\eqref{esbv},\eqref{epsdip}
\begin{equation}
\aver{\left(\frac{3}{2}T_a-\frac{1}{2}E_{\rm pair}-\frac{1}{3}E_{\rm rel}\right)\sigma_{\rm rec}v} = \frac{5\beta-11}{3}
\frac{\Gamma(2-\frac{\beta}{2}) }{2^{\frac{\beta+1}{2}}\sqrt{\pi}}\sigma_0
\left(\frac{\mu}{T_a}\right)^{\frac{\beta-1}{2}}T_a.
\label{esvdipA}
\end{equation}

It is convenient to pass from $T_a$ to new variable $\theta$:
\begin{equation}
T_a=\theta T^2/\bar {T}_{ay}.
\label{theta}
\end{equation}
Then Eq.(\ref{dT3}) reduces to the second equation of the system (\ref{system}). 
Hubble parameter at RD-stage is $H=\sqrt{\frac{4\pi^3g_{\epsilon}}{45}}\frac{T^2}{m_{\rm Pl}}$ with $m_{\rm Pl}$ being the Plank mass. At MD-stage it can be roughly given by $H(RD)\sqrt{T_{\rm RM}/T}$, where $T_{\rm RM}\sim 1$ eV is the temperature when RD$\rightarrow$MD transition occurs. The late $\Lambda$-dominated stage is not considered.

The system (\ref{system}) can be transformed to
\begin{equation}
\begin{cases}
\cfrac{dr}{dT}=D_s\cfrac{r^2}{\theta^{\frac{\beta-1}{2}}T^{\beta_s+1}}\\
\frac{d\theta}{dT}=-\gamma D_s\, \cfrac{r\,\theta^{\frac{3-\beta}{2}}}{T^{\beta_s+1}}.
\end{cases}
\label{system2}
\end{equation}
Here the following notations are introduced: index $s=$'R' or 'M' means RD- or MD- stage, 
$$
	\gamma=\frac{5\beta-11}{18},\quad \beta_{\rm R}=\beta-2,\quad \beta_{\rm M}=\beta-\frac{5}{2},
$$
$$
	D_{\rm R}=D_{\rm M}\sqrt{T_{\rm RM}}=\cfrac{\Gamma\left(2-\frac{\beta}{2}\right)g_s}{2^{\tfrac{\beta-3}{2}}\sqrt{45 g_{\epsilon}}}\;
	\sigma_0 \, m_\text{Pl}\left(\mu\bar{T}_{ay}\right)^{\tfrac{\beta-1}{2}}.
$$
To solve the system \eqref{system2}, one divides second equation by first, from where one gets independently on the stage
\begin{equation}
r(\theta)=r_0\, \theta^{-1/\gamma},
\label{rtheta}
\end{equation}
for initial conditions $\theta(T_0=\bar T_{ay})=1$, $r(T_0)=r_0$.
Being interested in $r(T)$ on MD-stage, we will put the solution of system
for RD-stage to be initial conditions for that for MD-stage. Substituting $r(\theta)$ of Eq.(\ref{rtheta}) in this manner into second equation of the system (\ref{system2}) yields
\begin{equation}
\theta(T)=\left\{ 1+ \frac{\bar{\gamma}}{\beta_{\rm R}} D_{\rm R} r_0 \left( \frac{1}{T_{\rm RM}^{\beta_{\rm R}}}-\frac{1}{T_0^{\beta_{\rm R}}}\right)
+\frac{\bar{\gamma}}{\beta_{\rm M}} D_{\rm M} r_0 \left( \frac{1}{T^{\beta_{\rm M}}}-\frac{1}{T_{\rm RM}^{\beta_{\rm M}}} \right) \right\}
^{\gamma/\bar{\gamma}},
\label{thetaTD}
\end{equation}
where $\bar{\gamma}=1+\gamma \frac{\beta-1}{2}$.
Function $\theta(T)$, with $\beta=14/5$, is very slowly growing with decrease of $T$ 
($\beta_{\rm R}=4/5$, $\beta_{\rm M}=3/10$, $\gamma=1/6$, $\bar{\gamma}=23/20$). 
At $T\ll T_{\rm RM}\ll T_0$, 
$\theta(T)\propto T^{-1/23}$.

Solution for $r(T)$ is given by Eqs.\eqref{rtheta} and \eqref{thetaTD}. 
At $T\ll T_{RM}\ll T_0$ (with $\beta_{\rm R,M}>0$)
\begin{equation}
r\approx \left\{ r_0^{\bar{\gamma}-1} \frac{\beta_{\rm M}}{\bar{\gamma}} \frac{T^{\beta_{\rm M}}}{D_{\rm M}} \right\}^{1/\bar{\gamma}}
\label{rapprox}
\end{equation}
Note, that a weak sensitivity of final density to its initial value ($r_0$) is obliged to thermal effects (second equation of \eqref{system}), and it vanishes totally if to ignore them ($\theta\equiv 1$, $\gamma=0$, $\bar{\gamma}= 1$).

\subsection*{Appendix C}

The action of the pair of mutually attracted particles looks like
\begin{equation}
S=\int\limits_{t_1}^{t_2}\left(\frac{\mu v^2}{2}+\frac{\alpha_y}{r}\right)dt.
\end{equation}
Here $v$ is the current velocity rather than initial one as defined above. In the region of interest 
we have $\frac{\mu v^2}{2}\sim \frac{\alpha_y}{r}$, whence $v\sim \sqrt{\frac{2\alpha_y}{\mu r}}$. One replaces $dt$ by $dr$
\begin{equation}
S\sim \int\limits_{r_1}^{r_2}2\frac{\alpha_y}{r}\frac{dr}{v}\sim \sqrt{\mu \alpha_y}\left(\sqrt{r_2}-\sqrt{r_1}\right).
\end{equation}
If we choose $r_1=R_b$ and $r_2=L_{\rm loss}+R_b\approx 10^{2/5} R_b$ then condition $S\gg 1$ gives Eq.\eqref{cl}.
Condition $R_b\gg a_B$ (and consequently Eq.\eqref{cl}) can be explicitly derived if we take  $r_1=a_B$ and $r_2=R_b$. On this interval the most of energy ($\sim I$) is lost.



\providecommand{\href}[2]{#2}\begingroup\raggedright
\endgroup

\end{document}